\documentclass{llncs}
\usepackage {amssymb,latexsym,graphicx,epic,eepic,stmaryrd}

\newcommand {\implies} {\ensuremath{\, \Rightarrow \,}}
\newcommand {\ltl}[1] {\ensuremath{\mathrm{LTL}(#1)}}
\newcommand {\ltll} {\ensuremath{\mathrm{LTL}}}
\newcommand {\tltl}[1] {\ensuremath{\mathrm{TLTL}(#1)}}
\newcommand {\tltll} {\ensuremath{\mathrm{TLTL}}}

\newcommand {\mitle}[1] {\ensuremath{\mathrm{MTL}(#1)}}
\newcommand {\mitlel} {\ensuremath{\mathrm{MTL}}}
\newcommand {\mitl}[1] {\ensuremath{\mathrm{MITL}(#1)}}
\newcommand {\mitll} {\ensuremath{\mathrm{MITL}}}
\newcommand {\mitlde}[1] {\ensuremath{\mathrm{MTL}^{\Fm}(#1)}}
\newcommand {\mitldel} {\ensuremath{\mathrm{MTL}^{\Fm}}}

\renewcommand {\O}{\ensuremath{O}}
\newcommand {\Ominus}{\ensuremath{O\!\!\!{\raisebox{1.25pt}{-}}\,}}
\newcommand {\U}{\ensuremath{U}}
\renewcommand {\S}{\ensuremath{S}}
\newcommand {\true} {\ensuremath{\top}}

\newcommand {\And} {\ensuremath{\wedge}}
\newcommand {\Or} {\ensuremath{\vee}}
\newcommand {\F} {\ensuremath{\mbox{\boldmath $\Diamond$}}}
\newcommand {\Fm} {\ensuremath{\Diamond}}
\newcommand {\Fp} {\ensuremath{\mbox{\boldmath
      $\Diamond\!\!\!{\raisebox{1.25pt}{-}}$ }}}
\newcommand {\Fpm} {\ensuremath{\Diamond\!\!\!{\raisebox{1.25pt}{-}}}}
\newcommand {\G}{\ensuremath{\Box}}
\newcommand {\notof} {\ensuremath{\neg}}

\newcommand {\nat}{\ensuremath{\mathbb{N}}}

\newcommand {\rnonneg}{\ensuremath{\mathbb{R}^{\geq 0}}}

\newcommand {\qnonneg}{\ensuremath{\mathbb{Q}^{\geq 0}}}
\newcommand {\tw}[1] {\ensuremath{\mathit{tw}(#1)}}

\newcommand {\twfn} {\ensuremath{\mathit{tw}}}

\newcommand {\lsymb}[1] {\ensuremath{L_{\mathit{sym}}(#1)}}

\newcommand {\two}[1] {\ensuremath{\mathit{T} #1^{\omega}}}

\newcommand {\edge} {\ensuremath{\stackrel{}{\longrightarrow}}}
\newcommand {\aedge}[1] {\ensuremath{\stackrel{#1}{\longrightarrow}}}

\newcommand {\union} {\ensuremath{\cup}}
\newcommand {\intersection} {\ensuremath{\cap}}
\newcommand {\intervals} {\ensuremath{\mathcal{I}_{\mathbb{Q}}}}

\newcommand {\mso}[1] {\ensuremath{\mathrm{MSO}(#1)}}
\newcommand {\msol} {\ensuremath{\mathrm{MSO}}}

\newcommand {\tmso}[1] {\mathrm{TMSO}{(#1)}}
\newcommand {\tmsol} {\ensuremath{\mathrm{TMSO}}}
\newcommand {\rtmso}[1] {\mathrm{rec\textrm{-}TMSO}{(#1)}}
\newcommand {\rtmsol} {\ensuremath{\mathrm{rec\textrm{-}TMSO}}}
\newcommand {\rtltl}[1] {\ensuremath{\mathrm{rec\textrm{-}TLTL(#1)}}}
\newcommand {\rtltll} {\ensuremath{\mathrm{rec\textrm{-}TLTL}}}

\newcommand {\ldist}[1] {\ensuremath{\vartriangleleft_{#1}}}
\newcommand {\rdist}[1] {\ensuremath{\vartriangleright_{#1}}}

\newcommand {\ttos}[1] {\ensuremath {\textit{t-s\/}(#1)}}
\newcommand {\ttosfn} {\ensuremath {\textit{t-s\/}}}
\newcommand {\stot}[1] {\ensuremath {\textit{s-t\/}(#1)}}
\newcommand {\stotfn} {\ensuremath {\textit{s-t\/}}}

\newcommand {\ivoc}[1] {\ensuremath{\mathit{ivoc}(#1)}}

\newcommand {\tfo}[1] {\ensuremath{\mathrm{TFO}}(#1)}
\newcommand {\tfol} {\ensuremath{\mathrm{TFO}}}
\newcommand {\fo}[1] {\ensuremath{\mathrm{FO}(#1)}}
\newcommand {\rtfo}[1] {\ensuremath{\mathrm{rec\textrm{-}TFO}}(#1)}
\newcommand {\rtfol} {\ensuremath{\mathrm{rec\textrm{-}TFO}}}

\newcommand {\sem}[1] {\ensuremath{\llbracket #1 \rrbracket}}
\newcommand {\Op} {\ensuremath{\mathit{Op}}}
\newcommand {\Rop} {\ensuremath{\mathit{Rop}}}
\newcommand {\pos}[1] {\ensuremath{\mathit{pos}(#1)}}
\newcommand {\idta} {IDA}
\newcommand {\ridta} {rec-IDA}
\newcommand {\fridta} {frec-IDA}
\newcommand {\fw}[1] {\ensuremath{\mathit{fw}(#1)}}
\newcommand {\fwfn} {\ensuremath{\mathit{fw}}}

\newenvironment {lemma*} {\noindent {\bf Lemma} \em} {\rm}

\begin{document}
\title{On timed automata with input-determined guards}
\author{Deepak D'Souza\inst{1} and Nicolas Tabareau\inst{2}}
\institute{Dept. of Computer Science \& Automation\\
Indian Institute of Science, Bangalore, India.\\
\email{deepakd@csa.iisc.ernet.in}
\and
\'Ecole Normale Superieure de Cachan, 
Cachan, France.\\
\email{Nicolas.Tabareau@dptmaths.ens-cachan.fr}}

\maketitle

\begin{abstract}
We consider a general notion of timed automata with
\emph{input-determined} guards and show that they admit a robust
logical framework along the lines of \cite{d03}, in terms of a monadic second order logic
characterisation and an expressively complete timed temporal
logic. 
We then generalise these automata using the notion of recursive
operators introduced by Henzinger, Raskin, and Schobbens \cite{hrs98}, and
show that they admit a similar logical framework.
These results hold in the ``pointwise'' semantics.
We finally use this framework to show that the
real-time logic MITL of Alur et al \cite{afh96} is expressively
complete with respect to an MSO corresponding to an appropriate
input-determined operator.

\end{abstract}

\section {Introduction}

The timed automata of Alur and Dill \cite{ad94} are a popular
model for describing timed behaviours.
While these automata have the plus point of being very expressive 
and having a decidable emptiness problem, they are 
neither determinisable nor closed under complementation.
This is a drawback from a couple of points of view.
Firstly, one cannot carry out model checking in the framework
where a system is modelled as a timed transition system
$\mathcal{T}$ and a specification of timed behaviours as a timed
automaton $\mathcal{A}$, and where one asks ``is
$L(\mathcal{T}) \subseteq L(\mathcal{A})$?''.
This would
normally involve \emph{complementing} $\mathcal{A}$ and then checking if
its intersection with $\mathcal{T}$ is non-empty.
One can get around this problem to some extent by using
determinisable specifications, or specifying directly the negation
of the required property.
A second reason why lack of closure properties may concern us is
that it precludes the existence of an unrestricted logical
characterisation of the class of languages accepted by timed
automata.
The existence of a monadic second order logic (MSO) characterisation of
a class of languages is a strong endorsement of the ``regularity''
of the class.
It also helps in identifying expressively complete
temporal logics, which are natural to use as specification
languages and have relatively efficient model checking algorithms.

The event clock automata of \cite{afh94} was one of the first
steps towards identifying a subclass of timed automata with the
required closure properties.
They were shown to be determinisable in \cite{afh94}, and later to
admit a robust logical framework in terms of an
MSO characterisation and an expressively complete timed temporal
logic \cite{d03}.
Similar results were shown in \cite{rs97}, \cite{hrs98} and
\cite{dt99}.
A common technique used in all these results was the idea
of ``implicit'' clocks, whose values are determined solely by
the timed word being read.
For example the event recording clock $x_a$ records the time since
the last $a$ action w.r.t.\@ the current position in a timed word,
and is thus implicitly reset with each $a$ action.
The truth of a guard over these clocks at a point in a timed
word is thus completely determined by the word itself, unlike in a
timed automaton where the value of a clock depends on 
the path taken in the automaton.

In this paper we generalise the notion of an implicit clock to that
of an \emph{input determined operator}.
An input determined operator $\Delta$ identifies
for a given timed word and position in it,
a set of intervals in which it is ``satisfied''.
The guard $I \in \Delta$ is then satisfied at a point in a timed
word if the set of intervals identified by $\Delta$ contains $I$.
For example, the event recording clock $x_a$ can be modelled as an
input determined operator $\ldist{a}$ which identifies at a given
point in a timed word, the (infinite) set of intervals containing
the distance to the last $a$ action.
The guard $(x_a \in I)$ now translates to $(I \in \ldist{a})$.
As an example to show that this framework is more general than
implicit clocks, consider the input determined operator $\F_a$
inspired by the Metric Temporal logic (MTL) of \cite{k90,ah93}.
This operator identifies the set of all intervals $I$ for
which there is a future occurrence of an $a$ at a distance which
lies in $I$.
The guard $I \in \F_a$ is now true iff there is a future occurrence
of an $a$ action, at a distance which lies in $I$.

Timed automata which use guards based on a set of input 
determined operators are
what we call \emph{input determined automata}.
We show that input determined automata form a robust class of
timed languages, in that they are (a) determinisable, (b)
effectively closed
under boolean operations, (c) admit a logical characterisation via
an unrestricted MSO, and (d) identify a
natural expressively complete timed temporal logic.

We then go over to a more expressive framework using the
idea of \emph{recursive} event clocks from \cite{hrs98}.
In the recursive version of our input determined operator, the
operators now expect a third parameter (apart from the timed word
and a position in it) which identifies a set of positions in the
timed word.
This argument could be (recursively) another input determined
automaton, or as is better illustrated, a temporal logic
formula $\theta$.
The formula $\theta$ naturally identifies a set of positions in a
timed word where the formula is satisfied.
Thus a recursive operator $\Delta$ along with the formula
$\theta$, written $\Delta_{\theta}$, behaves like an input
determined operator above, and the guard $I \in \Delta_{\theta}$
is true iff the set of intervals identified by $\Delta_{\theta}$
contains $I$.
These recursive input determined automata are also shown to admit
similar robust logical properties above.

We should be careful to point out here that, firstly, these results
hold in the \emph{pointwise} semantics, where formulas are
evaluated only at the ``action points'' in a timed word
(used e.g.\@ in \cite{w94}), and not at arbitrary points in between actions
in a timed word as allowed in the \emph{continuous} semantics
of \cite{afh96,hrs98}.
Secondly, we make no claims about the existence of
\emph{decision procedures} for these automata and logics.
In fact it can be seen the operator $\F_a$ above takes us out of
the class of timed automata as we can define the language of timed
sequences of $a$'s in which no two $a$'s are a distance 1 apart,
with a single state input determined automaton which has the guard
$\notof([1,1] \in \F_a)$.
Similar versions can be seen to have undecidable emptiness problems
and correspondingly undecidable logics \cite{ah93}.
Thus the contribution of this paper should be seen more in terms
of a general framework for displaying logical characterisations of
timed automata, and proving expressive completeness of temporal logics
related to these automata.
Many of the results along these lines from \cite{dt99,d03}
and some in the pointwise semantics from \cite{r99}
follow from the results in this paper.

As a new application of this framework, we provide an expressive
completeness result for MITL in the pointwise semantics, by
showing that it is expressively equivalent to the first order
fragment of an MSO based on recursive operators.
This answers an open question from \cite{r99}, apart from
identifying an interesting class of timed automata.

The techniques used in this paper essentially build on those from
\cite{dt99} and \cite{d03} which use the notion of \emph{proper}
symbolic alphabets and factor through the results of B\"uchi
\cite{b60} and Kamp \cite{k68}.
The idea of using recursive operators comes from \cite{hrs98}, who
show a variety of expressiveness results, including an expressive
completeness for MITL in the continuous semantics.
Their result for MITL is more interesting in that
it uses event-clock modalities, while we use essentially the same
modalities as MITL.
However, our MSO is more natural as unlike the MSO in \cite{hrs98}
it has unrestricted second order quantification.

\section{Input determined automata}
\label{section:idta}

We use $\nat$ to denote the set of natural numbers 
$\{0, 1, \ldots \}$, and $\rnonneg$ and $\qnonneg$ to
denote the set of non-negative reals and rationals
respectively.
The set of finite and infinite words over an
alphabet $A$ will be denoted by $A^{\ast}$ and $A^{\omega}$
respectively.
We use the notation $X \rightarrow Y$ to denote the set of
functions from $X$ to $Y$.

An \emph{(infinite) timed word} over an alphabet $\Sigma$ is an
element $\sigma$ of $(\Sigma \times \rnonneg)^\omega$ satisfying the
following conditions.
Let $\sigma = (a_0, t_0) (a_1, t_1) \cdots$.
Then:
\begin{enumerate}
\item
(\emph{monotonicity})
for each $i \in \nat$, $t_i \leq t_{i+1}$,
\item
(\emph{progressiveness})
for each $t \in \rnonneg$ there
exists $i \in \nat$ such that $t_i > t$.
\end{enumerate}
Let $\two{\Sigma}$ denote the set of 
infinite timed words over $\Sigma$.
Where convenient, we will use the representation
of $\sigma$ as $(\alpha, \tau)$ where 
$\alpha \in \Sigma^{\omega}$ and 
$\tau : \nat \rightarrow \rnonneg$ is a time sequence satisfying the
conditions above.

We will use rational bounded intervals
to specify timing constraints.
These intervals can be open or closed, and we allow
$\infty$ as an open right end.
These intervals denote a subset of reals in the usual manner --
for example $[2, \infty)$ denotes the set
$\{ t \in \rnonneg \ | \ 2 \leq t \}$.
The set of all such intervals is denoted $\intervals$.

Our input determined automata will use guards of the form ``$I \in
\Delta$'', where $I$ is an interval and $\Delta$ is an operator
which determines for a given timed word $\sigma$ and a position
$i$ in it, a set of intervals ``satisfying'' it at that point.
We then say that $\sigma$ at position $i$ satisfies the guard ``$I
\in \Delta$'' if $I$ belongs to the set of intervals
identified by $\Delta$.
By a ``position''
in the timed word we mean one of the ``action
points'' or instants given by the time-stamp sequence, and use
natural numbers $i$ (instead of the time $\tau(i)$) to denote these
positions.
More formally, an input determined operator $\Delta$ (w.r.t. the
alphabet $\Sigma$) has a
semantic function
$\sem{\Delta}: (\two{\Sigma} \times \nat) \rightarrow
2^{\intervals}$.
The guard $I \in \Delta$ is satisfied at position $i$ in $\sigma
\in \two{\Sigma}$
iff $I \in \sem{\Delta}(\sigma, i)$.

The transitions of our input determined automata are
labelled by symbolic actions of the form $(a,g)$ where $a$ is an action, 
and $g$ is a guard which is a boolean combination of 
atomic guards of the form $I \in \Delta$.
The set of guards over a finite set of input determined operators
$\Op$ is denoted by $\mathcal{G}(\Op)$ and given by the syntax
$g ::= \true \ | \ I \in \Delta \ | \ \notof g \ | \ g \Or g \ | \ g \And g$.
The satisfaction of a guard $g$ in a timed word $\sigma$ at position
$i$, written $\sigma,i \models g$, is given in the expected way: we have 
$\sigma, i \models \true$ always, $\sigma, i \models I \in \Delta$
as above, and the boolean operators $\notof$, $\Or$, and $\And$ 
interpreted as usual.

A \emph{symbolic alphabet} $\Gamma$ based on $(\Sigma, \Op)$ is a finite
subset of $\Sigma \times \mathcal{G}(\Op)$.
An infinite word $\gamma$ in $\Gamma^\omega$ specifies in a natural way a
subset of timed words $\tw{\gamma}$ defined as follows.
Let $\gamma(i) = (a_i, g_i)$ for each $i \in \nat$.
Let $\sigma \in \two{\Sigma}$ with $\sigma(i) = (b_i, t_i)$
for each $i \in \nat$.
Then $\sigma \in \tw{\gamma}$ iff for each $i \in
\nat$, $b_i = a_i$ and $\sigma, i \models g_i$.
We extend the map $\twfn$ to work on subsets of
$\Gamma^{\omega}$ in the natural way.
Thus, for $\widehat{L} \subseteq \Gamma^{\omega}$, we define
$\tw{\widehat{L}} = \bigcup_{\gamma \in
\widehat{L}} \tw{\gamma}$.
Finally, we denote the vocabulary of intervals mentioned in
$\Gamma$ by $\ivoc{\Gamma}$.

Recall that a B\"{u}chi automaton over an alphabet
$A$ is a structure
$\mathcal{A} = (Q, s, \edge, F)$
where $Q$ is a finite set of states,
$s \in Q$ is an initial state,
$\longrightarrow \subseteq
Q \times A \times Q$ is the transition relation, and
$F \subseteq Q$ is a set of accepting states.
Let $\alpha \in A^{\omega}$. 
A run of $\mathcal{A}$ over $\alpha$ is a map $\rho :
\nat \rightarrow Q$ which satisfies:
$\rho(0) = s$ and 
$\rho(i) \aedge{\alpha(i)} \rho(i+1)$ for every $i \in \nat$.
We say $\rho$ is an \emph{accepting} run of $\mathcal{A}$ on $\alpha$ if 
$\rho(i) \in F$ for infinitely many $i \in \nat$.
The set of words accepted by $\mathcal{A}$, denoted 
here as $\lsymb{\mathcal{A}}$ (for the ``symbolic'' language
accepted by $\mathcal{A}$), is defined to be
the set of words in $A^{\omega}$
on which $\mathcal{A}$ has an accepting run.

We are now in a position to define an input determined
automaton.
An \emph{input determined automaton} (\idta\ for short)
over an alphabet
$\Sigma$ and a set of operators $\Op$, is simply
a B\"uchi automaton over a symbolic alphabet based on 
$(\Sigma,\Op)$.
Viewed as a
B\"uchi automaton over a symbolic alphabet $\Gamma$, 
an input determined automaton $\mathcal{A}$ accepts
the language $\lsymb{\mathcal{A}} \subseteq
\Gamma^{\omega}$ which we call the symbolic
language accepted by $\mathcal{A}$.
However, we will be more interested in the timed language
accepted by $\mathcal{A}$: this is denoted $L(\mathcal{A})$
and is defined to be $\tw{\lsymb{\mathcal{A}}}$.

To give a concrete illustration of input determined automata,
we show how the event clock automata of \cite{afh94} can be
realized in the above framework.
Take $\Op$ to be the set of operators $\{\ldist{a}, \rdist{a} \ |
\ a \in \Sigma\}$, where the operators $\ldist{a}$ and $\rdist{a}$
essentially record the time since the last $a$ action, and the
time to the next $a$ action.
The operator $\ldist{a}$ (and similarly $\rdist{a}$) can be defined here
by setting $\sem{\ldist{a}}(\sigma, i)$ to be
\[
\{ I \in
\intervals \ | \ \exists j < i: \, \sigma(j) = a, \,
\tau(i) - \tau(j) \in I, 
\mathrm{\, and \, }
\forall k: \, j < k < i, \,\sigma(k) \neq a\}.
\]

As another example which we will use later in the paper,
consider the operator $\F_a$ related to MTL \cite{k90,ah93}.
The guard $\F_a \in I$ is meant to be true in a word $\sigma$ at
time $i$ iff there is a future
instant $j$ labelled $a$ and the distance to it lies in $I$ --
i.e. $\tau(j) - \tau(i) \in I$.
The guard $\Fp_a \in I$ makes a similar assertion about the
\emph{past} of $\sigma$ w.r.t. the current position.
An input determined automaton based on these operators can be
defined by taking
$\Op = \{\F_a, \Fp_a \ | \ a \in \Sigma\}$, and where, for example,
$\sem{\F_a}(\sigma, i) = \{ I \ | \ \exists j \geq i: \ \sigma(j)
= a, \mathrm{\ and \ } \tau(j) - \tau(i) \in I \}$.

We now want to show that the class of timed languages accepted by
input determined automata
(for a given choice of $\Sigma$ and $\Op$) is closed under
boolean operations.
The notion of a \emph{proper} symbolic alphabet will play an
important role here and subsequently.
A \emph{proper symbolic alphabet} based on $(\Sigma, \Op)$ is of
the form $\Gamma = \Sigma \times (\Op \rightarrow
2^{\mathcal{I}})$ where $\mathcal{I}$ is a finite subset of
$\intervals$.
An element of $\Gamma$ is thus of the form $(a,h)$, where
the set of intervals specified by $h(\Delta)$ is interpreted
as the \emph{exact} subset of intervals in 
$\ivoc{\Gamma}$ 
which are satisfied by $\Delta$.
This is formalised in the following definition of $\twfn_{\Gamma}$ for a
proper symbolic alphabet $\Gamma$.
Let $\gamma \in \Gamma^{\omega}$ with $\gamma(i) = (a_i, h_i)$.
Let $\sigma \in \two{\Sigma}$ with $\sigma(i) = (b_i, t_i)$.
Then $\sigma \in \twfn_{\Gamma}(\gamma)$ iff for each $i \in
\nat$: $b_i = a_i$ and for each $\Delta \in \Op$, 
$h_i(\Delta) = \sem{\Delta}(\sigma, i) \intersection
\ivoc{\Gamma}$.

Let $\Gamma$ be a proper symbolic alphabet based on $(\Sigma, \Op)$.
Then a B\"uchi automaton $\mathcal{A}$ 
over $\Gamma$, which we call a \emph{proper} \idta\ over $(\Sigma,
\Op)$, determines a timed language
over $\Sigma$ given by $\twfn_{\Gamma}(\lsymb{\mathcal{A}})$.

The class of timed languages defined by \idta's and proper \idta's
over $(\Sigma, \Op)$ coincide.
An \idta\ over a symbolic alphabet
$\Gamma$ can be converted to an equivalent one (in terms of the
timed language they define) over a proper symbolic alphabet
$\Gamma' = \Sigma \times (\Op \rightarrow 2^{\ivoc{\Gamma}})$.
Firstly, each transition label $(a,g)$ in $\Gamma$ can be written
in a disjunctive normal form $(c_1 \And \cdots \And c_k)$, with each
$c_i$ being a conjunction of literals $I \in \Delta$ or $\notof (I
\in \Delta)$.
Thus each transition labelled $(a,g)$ can be replaced by a set of
transitions labelled $(a,c_i)$, one for each $i$.
Now each transition labelled $(a,c)$, with $c$ a conjunct
guard,
can be replaced by a set of transitions $(a,h)$, one for each $h$
``consistent'' with $c$: i.e.\@ $h$ should satisfy the condition
that if $I \in \Delta$ is one of the conjuncts in
$c$ then $I \in h(\Delta)$, and if $\notof(I \in \Delta)$ 
is one of the conjuncts in $c$ then $I \not\in h(\Delta)$.
In the other direction, to go from a proper \idta\ to an \idta, 
a label $(a,h)$ of a proper symbolic
alphabet can be replaced by the guard 
\[
\bigwedge_{\Delta \in \Op} (\bigwedge_{I \in h(\Delta)} (I \in \Delta)
\ \And \ \bigwedge_{I \in \ivoc{\Gamma} - h(\Delta)} \notof(I \in \Delta)).
\]

The following property of proper symbolic alphabets will play a crucial
role.

\begin{lemma}
\label{lemma:proper1}
Let $\Gamma$ be a proper symbolic alphabet based on $\Sigma$.
Then for any $\sigma \in \two{\Sigma}$ there is a \emph{unique}
symbolic word $\gamma$ in $\Gamma^{\omega}$ such that $\sigma \in
\twfn_{\Gamma}(\gamma)$.
\end{lemma}

\begin{proof}
Let $\sigma(i) = (a_i, t_i)$.
Then the only possible symbolic word $\gamma$ we can use
must be given by $\gamma(i) = (a_i, h_i)$, where
$h_i(\Delta) = \sem{\Delta}(\sigma, i) \intersection
  \ivoc{\Gamma}$.
\qed
\end{proof}


In the light of lemma~\ref{lemma:proper1}, going from a symbolic
alphabet to a proper one can be viewed as a step towards
determinising the automaton with respect to its timed language.
From here one can simply use classical automata
theoretic techniques to determinise the automaton w.r.t. its
\emph{symbolic} language.
(Of course, since we deal with infinite words we will need to go
from a B\"uchi to 
a Muller or Rabin acceptance condition \cite{t90}).

\begin{theorem}
\label{thm:closure1}
The class of \idta's over $(\Sigma, \Op)$
are effectively closed under the
boolean operations of union, intersection, and complement.
\end{theorem}

\begin{proof}
It is sufficient to address union and complementation.
Given automata $\mathcal{A}$ and $\mathcal{B}$ over symbolic
alphabets $\Gamma$ and $\Lambda$ respectively, we can simply
construct an automaton over $\Gamma \union \Lambda$ 
which accepts the union of the two symbolic languages.
For complementing the timed language of $\mathcal{A}$, we can
go over to an equivalent proper \idta\ $\mathcal{A}'$ over a
 proper symbolic alphabet $\Gamma'$, and now simply
complement the symbolic language accepted by $\mathcal{A}'$ to get
an automaton $\mathcal{C}$.
It is easy to verify, using the uniqueness property of proper
alphabets given in Lemma~\ref{lemma:proper1}, 
that $L(\mathcal{C}) = \two{\Sigma} - L(\mathcal{A}')$.
In the constructions above we have made use of the 
closure properties of $\omega$-regular
languages \cite{t90}.
\qed
\end{proof}

\section{A logical characterisation of \idta's}
\label{section:tmso}

We now show that input determined automata admit a natural
characterisation via a timed MSO in the
spirit of \cite{b60}.
Recall that for an alphabet $A$, B\"uchi's monadic second
order logic (denoted here by $\mso{A}$) is given as follows:
\[
\varphi ::= Q_a(x) \ |\ x \in X \ | \ x < y \ | \ 
 \notof{\varphi} \ | \ (\varphi \Or \varphi)
\ | \ \exists x \varphi \ | \ \exists X \varphi. 
\]
The logic is interpreted over a word $\alpha \in A^{\omega}$,
along with an interpretation $\mathbb{I}$ which assigns individual
variables $x$ a position in $\alpha$ (i.e. an $i \in \nat$), and
to set variables $X$ a set of positions $S \subseteq \nat$.
The relation $<$ is interpreted as the usual ordering of natural
numbers, and the predicate $Q_a$ (one for each $a \in A$) as the
set of positions in $\alpha$ labelled $a$.

The formal semantics of the logic is given below.
For an interpretation $\mathbb{I}$ we use the notation
$\mathbb{I}[i/x]$ to denote the interpretation which sends $x$ to $i$
and agrees with $\mathbb{I}$ on all other variables.
Similarly, $\mathbb{I}[S/X]$ denotes the modification of $\mathbb{I}$
which maps the set variable $X$ to a subset $S$ of $\nat$.
Later we will also use the notation $[i/x]$ to denote the
interpretation with sends $x$ to $i$ when the rest of the
interpretation is irrelevant.

\[
\begin{array}{lll}
\alpha, \mathbb{I} \models Q_a(x) & \mathrm{\ iff \ } &
\alpha(\mathbb{I}(x)) = a.\\
\alpha, \mathbb{I} \models x \in X & \mathrm{\ iff \ } &
\mathbb{I}(x) \in \mathbb{I}(X). \\
\alpha, \mathbb{I} \models x < y & \mathrm{\ iff \ } &
\mathbb{I}(x) < \mathbb{I}(y). \\
\alpha, \mathbb{I} \models \exists x \varphi & \mathrm{\ iff \ } &
\mathrm{ there \ exists \ } i \in \nat \mathrm{ \
such \ that \ } \sigma,\mathbb{I}[i/x] \models \varphi.\\
\alpha, \mathbb{I} \models \exists X \varphi & \mathrm{\ iff \ } &
\mathrm{ there \ exists \ } S \subseteq \nat \mathrm{ \
such \ that \ } \sigma,\mathbb{I}[S/X] \models \varphi.
\end{array}
\]

For a sentence $\varphi$ (i.e. a formula without free variables) 
in $\mso{A}$ we set
\(
L(\varphi) = \{ \sigma \in A^{\omega} \ | \ \sigma \models
\varphi \}.
\)
B\"uchi's result then
states that a language $L \subseteq A^{\omega}$ 
is accepted by a B\"uchi automaton over $A$ 
iff $L = L(\varphi)$ for a sentence
$\varphi$ in $\mso{A}$.

We define a timed MSO called $\tmso{\Sigma,\Op}$, parameterised by
the alphabet $\Sigma$ and set of input determined operators $\Op$,
whose syntax is given by:
\[
\varphi ::= Q_a(x) \ | \ I \in \Delta(x) \ | \ 
x \in X \ | \ x < y \ | \ 
\notof{\varphi} \ | \ (\varphi \Or \varphi)
\ | \ \exists x \varphi \ | \ \exists X \varphi. 
\]
In the predicate ``$I \in \Delta(x)$'', $I$ is an interval in
$\intervals$, $\Delta \in \Op$, and $x$ is a variable.

The logic is interpreted in a similar manner to $\msol$, except
that models are now timed words over $\Sigma$.
In particular, for a timed word $\sigma = (\alpha, \tau)$, we have:
\[
\begin{array}{lll}
\sigma, \mathbb{I} \models Q_a(x) & \mathrm{iff} &
           \alpha(\mathbb{I}(x)) = a \\
\sigma, \mathbb{I} \models I \in \Delta(x) & \mathrm{iff} &
I \in \sem{\Delta}(\sigma, \mathbb{I}(x)).
\end{array}
\]

Given a sentence $\varphi$ in $\tmso{\Sigma}$ we define
$L(\varphi) = \{ \sigma \in \two{\Sigma} \ | \ \sigma
\models \varphi \}$.

\begin{theorem}
\label{thm:msoba1}
A timed language $L \subseteq \two{\Sigma}$ is accepted by an input
determined automaton over $(\Sigma, \Op)$ iff $L = L(\varphi)$ for
some sentence $\varphi$ in $\tmso{\Sigma, Op}$.
\end{theorem}

\begin{proof}
Given an \idta\ $\mathcal{A}$ 
over $(\Sigma, \Op)$ we can give
a $\tmsol$ sentence $\varphi$ which describes the existence of an accepting
run of $\mathcal{A}$ on a timed word.
Following \cite{t90}, for $\mathcal{A} = (Q, q_0, \edge, F)$ with
$Q = \{q_0, \ldots q_n\}$, we can take $\varphi$ to be the sentence
\begin{eqnarray*}
\lefteqn{\!\!\!\!\exists X_0 \cdots \exists X_n \ (\ 0 \in X_0 
             \ \And \ \displaystyle{\bigwedge_{i\neq j}} 
             \forall x (x \in X_i \implies \notof (x \in X_j)) } \\
(*)\ \ \ \ \      & & \And \ \forall x \displaystyle{\bigvee_{q_i \aedge{(a,g)} q_j}}
	         (x \in X_i \ \And \ (x+1) \in X_j \ \And
              Q_a(x) \And g') \\
         & & \And \ \displaystyle{\bigvee_{q_i \in F}} \forall x \exists y 
              (x < y \And y \in X_i)).
\end{eqnarray*}
Here $g'$ denotes the formula obtained by replacing each $I \in
\Delta$ in $g$ by $I \in \Delta(x)$.
Further, ``$0\in X_0$'' abbreviates $\forall x\, (\mathit{zero}(x) \implies
x \in X_0)$ where $\mathit{zero}(x)$ in turn stands for
$\notof\exists y (y < x)$.
Similarly $x+1 \in X_j$ can be expressed via 
$\forall y (\mathit{succ}_x(y) \implies y \in X_j)$, where
$\mathit{succ}_x(y)$ is the formula $x < y \ \And \ \notof\exists
z (x < z \ \And \ z < y)$.

In the converse direction we take the route used in \cite{d03} as
it will be useful in the sequel.
Let $\varphi$ be a formula in $\tmso{\Sigma, \Op}$, and let
$\Gamma$ be a \emph{proper} symbolic alphabet with the same
interval vocabulary as $\varphi$.
We give a way of translating $\varphi$
to a formula $\ttos{\varphi}$ in
$\mso{\Gamma}$ in such a way that the timed languages are
preserved.
The translation $\ttosfn$ is done with respect to $\Gamma$ and
simply replaces each occurrence of 
\[Q_a(x) \mathrm{\ \ by \ }
\bigvee_{(b,h) \in \Gamma,\ b=a} Q_{(b,h)}(x)
\mathrm{\ \ and \ \ } I \in \Delta(x) \mathrm{\ \ by \ }
\bigvee_{(a,h) \in \Gamma, \ I \in h(\Delta)} Q_{(a,h)}(x).
\]
The translation preserves the timed models of a formula $\varphi$ in the
following sense:
\begin{lemma}
Let $\sigma \in \two{\Sigma}$, $\gamma \in \Gamma^{\omega}$, and
$\sigma \in \twfn_{\Gamma}(\gamma)$.
Let $\mathbb{I}$ be an interpretation for variables.
Then $\sigma, \mathbb{I} \models \varphi$ iff $\gamma, \mathbb{I}
\models \ttos{\varphi}$.
\qed
\end{lemma}

The lemma is easy to prove using induction on the structure of the
formula $\varphi$ and making use of the properties of proper symbolic
alphabets.
From the lemma it immediately follows now that for a sentence
$\varphi$ in $\tmso{\Sigma,\Op}$, we have $L(\varphi) =
\twfn_{\Gamma}(L(\ttos{\varphi}))$, 
and this is the sense in which the translation
preserves timed languages.

We can now argue the converse direction of
Theorem~\ref{thm:msoba1}
using this translation and
factoring through B\"uchi's theorem.
Let $\varphi$ be a sentence in $\tmso{\Sigma, \Op}$ and let
$\widehat{\varphi} = \ttos{\varphi}$.
Then by B\"uchi's theorem we have an automaton $\mathcal{A}$ 
over $\Gamma$ which recognises exactly $L(\widehat{\varphi})$.
Thus $\mathcal{A}$ is our required proper \idta\, since
$L(\mathcal{A}) = \twfn_{\Gamma}(\lsymb{\mathcal{A}}) =
\twfn_{\Gamma}(L(\widehat{\varphi})) = L(\varphi)$.
\qed
\end{proof}

\section{An expressively complete timed $\ltll$}
\label{section:tltl}

In this section we identify a
natural, expressively complete, timed temporal logic based on
input determined operators.
The logic is denoted $\tltl{\Sigma,\Op}$, parameterised by the
alphabet $\Sigma$ and set of input determined operators $\Op$.
The formulas of $\tltl{\Sigma,\Op}$ are given by:
\[
\theta ::= \ a \ | \
I \in \Delta \ | \
\O \theta \ | \ 
\Ominus \theta \ | \ 
(\theta \U \theta) \ | \
(\theta \S \theta) \ | \
\notof \theta \ | \
(\theta \Or \theta).
\]
Here we require $a \in \Sigma$, $I \in \intervals$, and $\Delta \in \Op$.
The models for $\tltl{\Sigma, \Op}$ formulas are
timed words over $\Sigma$.
Let $\sigma \in \two{\Sigma}$, with $\sigma = (\alpha, \tau)$,
and let $i \in \nat $.
Then the satisfaction relation $\sigma, i \models \varphi$
is given by
\[
\begin{array}{lll}
\sigma, i \models a & \mathrm{\ iff \ } & \alpha(i) = a \\
\sigma, i \models I \in \Delta & \mathrm{\ iff \ } &
     I \in \sem{\Delta}(\sigma, i) \\
\sigma, i \models \O \theta & \mathrm{\ iff \ } & 
     \sigma, i+1 \models \theta \\
\sigma, i \models \Ominus \theta & \mathrm{\ iff \ } & 
     i > 0 \mathrm{\ and \ } \sigma, i-1 \models \theta \\
\sigma, i \models \theta \U \eta & \mathrm{\ iff \ } &
\exists k \geq i:\, \sigma, k \models \eta \mathrm{\ and \ }
\forall j:\, i \leq j < k, \ \sigma, j \models \theta \\
\sigma, i \models \theta \S \eta & \mathrm{\ iff\ } &
\exists k < i:\, \sigma, k \models \eta \mathrm{\ and \ }
\forall j:\, k < j \leq i, \ \sigma, j \models \theta \\
\end{array}
\]
We define $L(\theta) = 
\{ \sigma \in \two{\Sigma} \ | \ \sigma, 0 \models \varphi \}$.

Let us denote by $\tfo{\Sigma, \Op}$ the first-order fragment of
$\tmso{\Sigma, \Op}$ (i.e. the fragment we get by disallowing
quantification over set variables).
The logics $\tltll$ and $\tfol$ are \emph{expressively equivalent}
in the following sense:

\begin{theorem}
\label{thm:ltl1}
A timed language $L \subseteq \two{\Sigma}$ is definable
by a $\tltl{\Sigma, \Op}$ formula $\theta$ iff
it is definable by a sentence $\varphi$ in $\tfo{\Sigma,\Op}$.
\end{theorem}

\begin{proof}
Given a $\tltl{\Sigma, \Op}$ formula $\theta$ we can associate an
$\tfo{\Sigma, \Op}$ formula $\varphi$ which has a single free variable
$x$, and satisfies the property that $\sigma, i \models \theta$ iff 
$\sigma, [i/x] \models \varphi$.
This can be done in a straightforward inductive manner as follows.
For the atomic formulas $a$ and $I \in \Delta$ we can take $\varphi$ to be 
$Q_a(x)$ and $I \in \Delta(x)$ respectively.
In the inductive step, assuming we have already translated $\theta$
and $\eta$ into $\varphi$ and $\psi$ respectively, we can translate
$\theta \U \eta$ into 
\[
\exists y (x \leq y \And \psi[y/x] \And \forall z ((x \leq z \And z \leq
y) \implies \varphi[z/x])).
\]
Here $\psi[y/x]$ denotes the standard renaming of the free variable
$x$ to $y$ in $\psi$.
The remaining modalities are handled in a similar way, and we can
verify that if $\varphi$ is the above translation of $\theta$
then $\sigma, i \models \theta$ iff
$\sigma, [i/x] \models \varphi$.
It also follows that $\sigma,0$ satisfies $\theta$ iff
$\sigma$ satisfies the sentence $\varphi_0$ given by
$\forall x (\mathit{zero}(x) \implies \varphi)$.
Hence we have that $L(\theta) = L(\varphi_0)$.

In the converse direction a more transparent proof is
obtained by factoring through Kamp's result
for classical LTL.
Recall that the syntax of $\ltl{A}$ is given by:
\[
\theta ::= \ a \ | \
\O \theta \ | \
\Ominus \theta \ | \
(\theta \U \theta) \ | \
(\theta \S \theta) \ | \
\notof{\theta} \ | \
(\theta \Or \theta)
\]
where $a \in A$.
The semantics is given in a similar manner to $\tltll$, except that
models are words in $A^{\omega}$.
In particular the satisfaction relation $\alpha, i \models \theta$
for the atomic formula $a$ is given by:
$\sigma, i \models a$ iff $\alpha(i) = a$.
Let $\fo{A}$ denote the first-order fragment of $\mso{A}$.
Then the result due to Kamp \cite {k68} states that:
\begin{theorem}[\cite{k68}]
\label{thm:kamp1}
$\ltl{A}$ is expressively equivalent to $\fo{A}$.
\qed
\end{theorem}

Consider now a proper symbolic alphabet $\Gamma$ based on $(\Sigma, \Op)$.
We can define a timed language preserving 
translation of an $\ltl{\Gamma}$ formula
$\widehat{\theta}$ to a formula $\stot{\widehat{\theta}}$ in
$\tltl{\Sigma, \Op}$.
In the translation $\stotfn$ we replace subformulas $(a,h)$ by
\[
a \And \bigwedge_{\Delta \in \Op}(\bigwedge_{I \in h(\Delta)} (I \in \Delta)
\ \And \ \bigwedge_{I \in \ivoc{\Gamma} - h(\Delta)} 
                              \notof{(I \in \Delta)}).
\]
It is easy to argue along the lines of Lemma~\ref{lemma:proper1} that
\begin{lemma}
\label{lemma:stot-ltl}
Let $\sigma \in \two{\Sigma}$ and $\gamma \in \Gamma^{\omega}$ with
$\sigma \in \twfn_{\Gamma}(\gamma)$.
Then $\sigma, i \models \stot{\widehat{\theta}}$ iff $\gamma, i \models
\widehat{\theta}$.
\qed
\end{lemma}
Hence we have $L(\stot{\widehat{\theta}}) = \twfn_{\Gamma}(L(\widehat{\theta}))$.

We can now translate a sentence $\varphi$ in $\tfo{\Sigma, \Op}$ to
an equivalent $\tltl{\Sigma, \Op}$ formula $\theta$ as follows.
Let $\Gamma$ be the proper symbolic alphabet based on $(\Sigma, \Op)$
with the same interval vocabulary as $\varphi$.
Let $\widehat{\varphi}$ be the $\fo{\Gamma}$ formula $\ttos{\varphi}$.
Note that the translation $\stotfn$ preserves first-orderness and
hence $\widehat{\varphi}$ belongs to $\fo{\Gamma}$.
Now by Theorem~\ref{thm:kamp1}, we have a formula $\widehat{\theta}$ in
$\ltl{\Gamma}$ which is equivalent to $\widehat{\varphi}$.
We now use the translation $\ttosfn$ on the formula $\widehat{\theta}$
to get a $\tltl{\Sigma, \Op}$ formula $\theta$.
$\theta$ is our required $\tltl{\Sigma, \Op}$ formula.
Observe that firstly
$L(\theta) = \twfn_{\Gamma}(L(\widehat{\theta}))$ by the property of the
translation $\stotfn$.
Next, by Kamp's theorem we have that $L(\widehat{\theta}) =
L(\widehat{\varphi})$ and hence
$\twfn_{\Gamma}(L(\widehat{\theta})) =
\twfn_{\Gamma}(L(\widehat{\varphi}))$.
But by the property of the translation $\ttosfn$ applied to
$\varphi$, we have 
$\twfn_{\Gamma}(L(\widehat{\varphi})) = L(\varphi)$, and hence we can
conclude that $L(\varphi) = L(\theta)$.
This completes the proof of Theorem~\ref{thm:ltl1}.
\qed
\end{proof}

We point out here that the past temporal operators of $\Ominus$
(``previous'') and $\S$ (``since'') can be dropped from
our logic without affecting the
expressiveness of the logic.
This follows since it is shown in \cite{gpss80} that
Theorem~\ref{thm:kamp1} holds for the future fragment of $\ltll$.
The reason we retain the past operators is because they are needed
when we consider a recursive version of the logic in
Section~\ref{section:rltl}.

\section{Recursive input determined automata}

We now consider ``recursive'' input determined operators.
The main motivation is to increase the expressive power of our
automata, as well as to characterise the expressiveness of
recursive temporal logics which occur naturally in the real-time
setting.

To introduce recursion in our operators, we need to consider
\emph{parameterised} (or \emph{recursive}) input determined operators.
These operators, which we continue to denote by $\Delta$,
have a semantic function $\sem{\Delta} : (2^{\nat} \times \two{\Sigma}
\times \nat) \rightarrow 2^{\intervals}$, whose first argument
is a subset of positions $X$.
Thus $\Delta$ with the parameter $X$ determines 
an input determined operator of
the type introduced earlier, whose semantic function is given by
the map $(\sigma, i) \mapsto \sem{\Delta}(X, \sigma, i)$.
The set of positions $X$ will typically be specified by a
temporal logic formula or a  ``floating'' automaton,
in the sense
that given a timed word $\sigma$, the formula (resp. automaton) will
identify a set of positions in $\sigma$ where the formula is
satisfied (resp. automaton accepts).
These ideas will soon be made more precise.

We first recall the idea of a ``floating'' automaton introduced in
\cite{hrs98}.
These are automata which accept pairs of the form $(\sigma, i)$
with $\sigma$ a timed word, and $i$ a position (i.e.\@ $i \in
\nat)$.
We will represent a ``floating'' word $(\sigma,i)$ as a timed
word over $\Sigma \times \{0,1\}$.
Thus a timed word $\nu$ over $\Sigma \times \{0,1\}$
represents the floating word $(\sigma, i)$, iff
$\nu = (\alpha, \beta, \tau)$, with $\beta \in \{0,1\}^{\omega}$ with a
\emph{single} $1$ in the $i$-th position, 
and $\sigma = (\alpha, \tau)$.
We use $\fwfn$ to denote the (partial) map which given a timed word $\nu$
over $\Sigma \times \{0,1\}$ returns the floating word $(\sigma,
i)$ corresponding to $\nu$, and extend it to apply to timed languages
over $\Sigma \times \{0,1\}$ in the natural way.

Let $\Op$ be a set of input determined operators w.r.t. $\Sigma$.
Then a \emph{floating \idta} over $(\Sigma, \Op)$ is an \idta\
over $(\Sigma \times \{0,1\}, \Op')$, where the set of operators
$\Op'$ w.r.t. $\Sigma \times \{0,1\}$ is defined to be $\{ \Delta' \
| \ \Delta \in \Op \}$, with the semantics
\[
\sem{\Delta'}(\sigma', i) = \sem{\Delta}(\sigma, i),
\]
where $\sigma'$ is a timed word over $\Sigma \times \{0,1\}$,
with $\sigma' = (\alpha, \beta, \tau)$ and $\sigma = (\alpha,
 \tau)$.
Thus the operator $\Delta'$ simply ignores the $\{0,1\}$ component
of $\sigma'$ and behaves like $\Delta$ on the $\Sigma$ component.
A floating \idta\ $\mathcal{B}$ accepts the floating
timed language $L^f(\mathcal{B}) = \fw{L(\mathcal{B})}$.

We now give a more precise definition of recursive input
determined automata, denoted \ridta, and their floating
counterparts \fridta.
Let $\Rop$ be a finite set of recursive input determined
operators.
Then the class of \ridta's over $(\Sigma, \Rop)$, and the timed
languages they accept, are defined as follows.
\begin{itemize}
\item
Every \idta\ $\mathcal{A}$ over $\Sigma$ that uses only the guard
$\true$ is a \ridta\ over $(\Sigma,\Rop)$, and
accepts the timed language $L(\mathcal{A})$.

Similarly, every floating \idta\ $\mathcal{B}$ over $\Sigma$ which uses
only the guard $\true$ is
a \fridta\ over $(\Sigma, \Rop)$, and accepts the floating language
$L^f(\mathcal{B})$.

\item
Let $C$ be a finite collection of \fridta's over $(\Sigma, \Rop)$.
Let $\Op$ be the set of input determined operators
$\{\Delta_{\mathcal{B}} \ | \ \Delta \in \Rop, \ \mathcal{B} \in C
\}$, where the semantic function of each $\Delta_{\mathcal{B}}$ is
given as follows.
Let $\pos{\sigma, \mathcal{B}}$ denote the set of positions $i$
such that $(\sigma, i) \in L^f(\mathcal{B})$.
Then $\sem{\Delta_{\mathcal{B}}}(\sigma, i) =
\sem{\Delta}(\pos{\sigma, \mathcal{B}}, \sigma, i)$.

Then any \idta\ $\mathcal{A}$ over $(\Sigma, \Op)$
is a \ridta\ over $(\Sigma, \Rop)$, and accepts the timed
language $L(\mathcal{A})$ (defined in 
Section~\ref{section:idta}).

Similarly every floating \idta\ $\mathcal{B}$ 
over $(\Sigma, \Op)$ 
is a \fridta\ over $(\Sigma, \Rop)$, and
accepts the floating language $L^f(\mathcal{B})$.
\end{itemize}

Recursive automata fall into a natural ``level'' based on the
level of nesting of operators they use.
A \ridta\ is of \emph{level} 0 if the only guard it uses is
$\true$.
Similarly a \fridta\ is of level 0, if the only guard it uses is
$\true$.
A \ridta\ is of \emph{level} (i+1) if it uses an operator
$\Delta_{\mathcal{B}}$, with $\Delta \in \Rop$
and $\mathcal{B}$ a \fridta\ of level $i$,
and no operator $\Delta'_{\mathcal{C}}$ with $\Delta' \in \Rop$ and
$\mathcal{C}$ of level greater than $i$.
A similar definition of level applies to \fridta's.

As an example consider the level 1 \ridta\ $\mathcal{A}$ over the alphabet
$\{a,b\}$ below.
The floating automaton $\mathcal{B}$ accepts a floating word
$(\sigma, i)$ iff the position $i$ is labelled $b$ and the
previous and next positions are labelled $a$.
The recursive input determined operator $\F$ is defined formally
in Sec.~\ref{section:mitl}.
The \ridta\ $\mathcal{A}$ thus recognises the set of timed words
$\sigma$ over $\{a,b\}$ which begin with an $a$ and have an
occurrence of $b$ -- with $a$'s on its left and right -- exactly 1
time unit later.
\begin{center}
  \includegraphics[width=\textwidth]{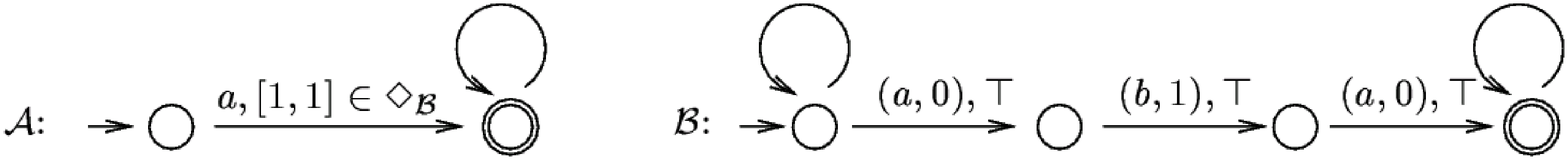}
\end{center}
\begin{theorem}
\label{thm:closure2}
The class of \ridta's over $(\Sigma, \Rop)$ is closed under boolean
operations.
In fact, for each $i$, the class of level $i$ \ridta's is closed
under boolean operations.
\end{theorem}

\begin{proof}
Let $\mathcal{A}$ and $\mathcal{A}'$ be two \ridta's of level $i$.
Let $\Op$ be the union of operators used in $\mathcal{A}$ and
$\mathcal{A}'$.
Then both $\mathcal{A}$ and $\mathcal{A}'$ are \idta's
over $(\Sigma, \Op)$, and hence by
Theorem~\ref{thm:closure1} there exists an \idta\ 
$\mathcal{B}$ over $(\Sigma, \Op)$ which accepts
$L(\mathcal{A}) \union L(\mathcal{A}')$.
Similarly there exists an \idta\ $\mathcal{C}$ over $(\Sigma, \Op)$,
which accepts the
language $\two{\Sigma} - L(\mathcal{A})$.
Notice that $\mathcal{B}$ and $\mathcal{C}$ use the same set of
operators $\Op$, and hence are also
level $i$ automata.
\qed
\end{proof}

We note that \idta's over $(\Sigma, \Op)$ are a special case of level
1 \ridta's over $(\Sigma, \Rop)$, where the set of recursive
operators $\Rop$ is taken to be $\{ \Delta' \ | \, \Delta \in
\Op\}$ with $\sem{\Delta'}(X,\sigma,i) = 
\sem{\Delta}(\sigma,i)$.
Thus each guard $I \in \Delta$ in an \idta\ over $(\Sigma, \Op)$
can be replaced by the guard $I \in
\Delta'_{\mathcal{B}}$, for any ``dummy'' level 0
\fridta\ $\mathcal{B}$.

\section{MSO characterisation of \ridta's}
\label{section:mso2}

We now introduce a recursive version of $\tmsol$ which will
characterise the class of timed languages defined by \ridta's.
The logic is parameterised by an alphabet $\Sigma$ and set of
recursive input determined operators $\Rop$, and denoted
$\rtmso{\Sigma, \Rop}$.
The syntax of the logic is given by
\[
\varphi ::= Q_a(x) \ | \ I \in \Delta_{\psi}(x) \ | \ 
x \in X \ | \ x < y \ | \ 
\notof{\varphi} \ | \ (\varphi \Or \varphi)
\ | \ \exists x \varphi \ | \ \exists X \varphi.
\]
In the predicate $I \in \Delta_{\psi}(x)$, 
we have $I \in \intervals$, $\Delta
\in \Rop$, and $\psi$ a $\rtmso{\Sigma, \Rop}$ formula with a
single free variable $z$.

The logic is interpreted over timed words in $\two{\Sigma}$.
Its semantics is similar to $\tmsol$ except for the predicate ``$I
\in \Delta_{\psi}(x)$'' which is defined inductively as follows.
If $\psi$ is a formula which uses no $\Delta$ predicates, then the
satisfaction relation $\sigma, \mathbb{I} \models \psi$ is defined
as for $\tmsol$.
Inductively, assuming the semantics of $\psi$ has already been
defined, $\Delta_{\psi}$ is interpreted as an input determined
operator as follows.
Let $\pos{\sigma, \psi}$ denote the set of interpretations for $z$
that make $\psi$ true in the timed word $\sigma$ --
i.e. $\pos{\sigma, \psi} = \{ i \ | \ \sigma, [i/z] \models
\psi\}$.
Then
\[
\sem{\Delta_{\psi}}(\sigma, i) = 
\sem{\Delta}(\pos{\sigma, \psi}, \sigma, i).
\]
Thus we have
\[
\sigma, \mathbb{I} \models I \in \Delta_{\psi}(x) \mathrm{\ iff \
} I \in \sem{\Delta}(\pos{\sigma, \psi}, \sigma, \mathbb{I}(x)).
\]

Note that the variable $z$, which is free in $\psi$, 
is \emph{not} free in the formula $I
\in \Delta_{\psi}(x)$.
A sentence $\varphi$ in $\rtmso{\Sigma, \Rop}$ defines the
language $L(\varphi) = \{ \sigma \models \varphi \}$, and
a $\rtmso{\Sigma, \Rop}$ formula $\psi$ with one free variable $z$
defines a floating language $L^f(\psi) = \{ \sigma, i \ | \
\sigma, [i/z] \models \psi \}$.

We note that each $\rtmso{\Sigma, \Rop}$ formula $\varphi$ 
can be viewed
as a $\tmso{\Sigma, \Op}$ formula, for a suitably defined set of
input determined operators $\Op$.
We say an operator $\Delta_{\psi}$ has a \emph{top-level}
occurrence in $\varphi$ if there is an occurrence of $\Delta_{\psi}$
in $\varphi$ which is \emph{not} in the scope of any
$\Delta'$ operator.
We can now take $\Op$ to be the set of all top-level operators
$\Delta_\psi$ in $\varphi$.

Analogous to the notion of level for \ridta's we can define the
\emph{level} of an $\rtmsol$ formula $\varphi$.
The level of $\varphi$ is 0, if $\varphi$ uses no $\Delta$
predicates.
$\varphi$ has level $i+1$ if it uses a predicate of the form $I
\in \Delta_{\psi}(x)$ with $\psi$ a level $i$ formula, and
\emph{no} predicate of the form $I \in \Delta'_{\phi}(x)$ with
$\phi$ of level greater than $i$.

As an example the level 1 sentence $\varphi$ below defines the same timed
language as the level 1 \ridta\ $\mathcal{A}$ defined in
Section~\ref{section:idta}.
We can take $\varphi$ to be
\(
\forall x (\mathit{zero}(x) \implies (Q_a(x)
\And ([1,1] \in \F_{\psi}(x)))),
\)
where $\psi$ is the level 0 formula
$Q_b(z) \And Q_a(z-1) \And Q_a(z+1)$.

\begin{theorem}
\label{thm:msoba2}
$L \subseteq \two{\Sigma}$ is accepted by a \ridta\ over $(\Sigma,
  \Rop)$ iff $L$ is definable by a $\rtmso{\Sigma, \Rop}$ sentence.
\end{theorem}

In fact, we will show that for each $i$, the class of \ridta's of level $i$
correspond to the sentences of $\rtmso{\Sigma, \Rop}$ of level
$i$.
But first it will be useful to state a
characterisation of floating languages along the lines of
Theorem~\ref{thm:msoba1}.

\begin{theorem}
\label{thm:msobaf1}
Let $L$ be a a floating language over $\Sigma$.
Then $L = L^f(\mathcal{B})$ for some floating \idta\ over
$(\Sigma, \Op)$ iff $L = L^f(\psi)$, for some 
$\tmso{\Sigma, \Op}$ formula $\psi$ with one free variable.
\end{theorem}

\begin{proof}
Let $\mathcal{B}$ be a floating \idta\ over $(\Sigma, \Op)$.
Keeping in mind that $\mathcal{B}$ runs over the alphabet $\Sigma
\times \{0,1\}$, we define a formula $\psi$ with one free variable
$z$ as follows.
$\psi$ is the formula $\varphi$ given in the proof
of Theorem~\ref{thm:msoba1}, except for the clause (*) which we
replace by

\begin{eqnarray*}
              & & \And \ \forall x ((x = z) \implies
\displaystyle{\bigvee_{q_i \aedge{((a,1),g)} q_j}}
	         (x \in X_i \ \And \ (x+1) \in X_j \ \And
                  Q_a(x) \And g') \\
                  & & \ \ \ \ \ \ \ \ \And \ (x \neq z) \implies
\displaystyle{\bigvee_{q_i \aedge{((a,0),g)} q_j}}
	         (x \in X_i \ \And \ (x+1) \in X_j \ \And
                  Q_a(x) \And g')).
\end{eqnarray*}
The formula $\psi$ satisfies $(\sigma,i) \in L^f(\mathcal{B})$ iff
$\sigma, [i/z] \models \psi$.

In the converse direction, let $\varphi(m,n)$ denote a
$\tmso{\Sigma, \Op}$ formula with free variables $x_1,\ldots,x_m,
X_1 \ldots X_n$.
An interpretation $\mathbb{I}$ for these variables is encoded
(along with $\sigma$) as a timed word over $\Sigma \times
\{0,1\}^{m+n}$.
We extend the definition of a floating \idta\ to an \idta\ which
works over such an alphabet, where, in particular, the $\Delta$ operators
apply only to the $\Sigma$ component of the timed word.
Then we can inductively associate with $\varphi(m,n)$ 
a floating \idta\ $\mathcal{B}$
over $\Sigma \times \{0,1\}$ such that $L^f(\mathcal{B}) =
L^f(\varphi)$.
In the inductive step for $\exists X_n(\varphi(m,n))$ we make use
of the fact that the class of languages accepted by floating
\idta's over $(\Sigma, \Op)$ are closed under the restricted
renaming operation required in this case.
The reader is referred to \cite{d03} for a similar argument.
\qed
\end{proof}

Returning now to the proof of Theorem~\ref{thm:msoba2}, we use
induction on the level of automata and formulas to argue that
\begin{enumerate}
\item
$L \subseteq \two{\Sigma}$ is accepted by a level $i$ \ridta\ over
  $(\Sigma, \Rop)$ iff $L$ is definable by a level $i$
  $\rtmso{\Sigma, \Rop}$ sentence $\varphi$. And
\item
\label{item:float}
A floating language $L$ over $\Sigma$
is accepted by a level $i$ \fridta\ over
  $(\Sigma, \Rop)$ iff $L$ is definable by a level $i$
  $\rtmso{\Sigma, \Rop}$ formula $\psi$ with one free variable.
\end{enumerate}

For the base case we consider level 0 automata and sentences.
Since level 0 automata only make use of the guard $\true$, they
are simply B\"uchi automata over $\Sigma$.
Similarly, level 0 sentences don't use any $\Delta$ predicates and
hence they are simply $\mso{\Sigma}$ sentences.
By B\"uchi's theorem, we have that level 0 automata and
sentences are expressively equivalent.

For the base case for the second part of the claim, given a level
0 floating automaton $\mathcal{B}$ we can apply the construction
in the proof of Theorem~\ref{thm:msobaf1} to get a $\tmso{\Sigma}$
formula $\psi$ with one free variable.
Since the construction preserves the guards used, $\psi$ has no
$\Delta$ operators, and hence is a level 0 $\rtmso{\Sigma, \Rop}$
formula.
Conversely, for a level 0 formula $\psi$ we can apply the
construction of Theorem~\ref{thm:msobaf1} to obtain a floating
automaton $\mathcal{B}$ such that $L^f(\mathcal{B}) = L^f(\psi)$.
The construction preserves the $\Delta$ operators used,
and hence $\mathcal{B}$ is a level 0 automaton.

Turning now to the induction step, let $\mathcal{A}$ be a level
$i+1$ automaton over $(\Sigma, \Rop)$.
Let $\Op$ be the set of top-level $\Delta$ operators in
$\mathcal{A}$.
Now since $\mathcal{A}$ is an \idta\  over $(\Sigma, \Op)$, by
Theorem~\ref{thm:msoba1}, we have a $\tmso{\Sigma, \Op}$ sentence
  $\varphi$ such that $L(\mathcal{A}) = L(\varphi)$.
Now for each $\Delta_{\mathcal{B}}$ in $\Op$, $\mathcal{B}$ is of
level $i$ or lower, and by our induction hypothesis there is a
corresponding $\rtmso{\Sigma, \Rop}$ formula
$\psi$ with one free variable, of the same level as $\mathcal{B}$,
with $L^f(\mathcal{B}) = L^f(\psi)$.
Hence for each $\Delta_{\mathcal{B}}$ we have a
semantically equivalent operator $\Delta_{\psi}$.
This is because $L^f(\mathcal{B}) = L^f(\psi)$, which implies 
$\pos{\sigma, \mathcal{B}} = \pos{\sigma, \psi}$, which in turn
implies $\sem{\Delta_{\mathcal{B}}} = \sem{\Delta_{\psi}}$.
We can now simply replace each occurrence of
$\Delta_{\mathcal{B}}$ in $\varphi$ to get an equivalent sentence
$\varphi'$ which is in $\rtmso{\Sigma, \Rop}$.
Further, by construction it follows that $\varphi'$ is also of
level $i+1$.

Conversely, let $\varphi$ be a level $i+1$ sentence in
$\rtmso{\Sigma, \Rop}$.
Let $\Op$ be the set of top level $\Delta$ operators in $\varphi$.
Then $\varphi$ is a $\tmso{\Sigma, \Op}$ sentence, and hence by
Theorem~\ref{thm:msoba1} we have an equivalent input determined
automaton $\mathcal{A}$ over $(\Sigma, \Op)$.
Once again, for each $\Delta_{\psi}$ in $\Op$, the formula $\psi$ is of level
$i$ or lower, and hence by induction hypothesis we have a 
\fridta\ $\mathcal{B}$ over $(\Sigma, \Rop)$, of the same level as
$\psi$, and accepting the same floating language.
The operators $\Delta_{\psi}$ and $\Delta_{\mathcal{B}}$ are now
equivalent, and we can replace each $\Delta_{\psi}$ in
$\mathcal{A}$ by the corresponding $\Delta_{\mathcal{B}}$ to get
a language equivalent input determined automaton.
This automaton is now the required level $i+1$ \ridta\ over
$(\Sigma, \Rop)$ which accepts the same language as $L(\varphi)$.

The induction step for part~\ref{item:float} is proved
similarly, making use of Theorem~\ref{thm:msobaf1} and the
induction hypothesis.
This completes the proof of Theorem~\ref{thm:msoba2}.
\qed

\section{Expressive completeness of \rtltll}
\label{section:rltl}

We now define a recursive timed temporal logic along the lines of
\cite{hrs98}.
The logic is similar to the logic $\tltll$ defined in
Sec.~\ref{section:tltl}.
It is parameterised by an alphabet $\Sigma$ and a set of
recursive input determined operators $\Rop$, and denoted
$\rtltl{\Sigma, \Rop}$.
The syntax of the logic is given by
\[
\theta ::= \ a \ | \
I \in \Delta_{\theta} \ | \
\O \theta \ | \ 
\Ominus \theta \ | \ 
(\theta \U \theta) \ | \
(\theta \S \theta) \ | \
\notof \theta \ | \
(\theta \Or \theta),
\]
where $a \in \Sigma$, and $\Delta \in \Rop$.

The logic is interpreted over timed words in a similar manner to
$\tltll$.
The predicate $I \in \Delta_{\theta}$ is interpreted as follows.
If $\theta$ does not use a $\Delta$ predicate, then the
satisfaction relation $\sigma, i \models \theta$ is defined as for
$\tltll$.
Inductively assuming the semantics of a $\rtltl{\Sigma, \Rop}$
formula $\theta$ has been defined, 
and setting
$\pos{\sigma, \theta} = \{ i \in \nat \ | \ \sigma, i \models
\theta \}$,
the operator $\Delta_{\theta}$
is interpreted as an input determined operator with the semantic
function
\[
\sem{\Delta_{\theta}}(\sigma, i) =
\sem{\Delta}(\pos{\sigma, \theta}, \sigma, i).
\]
The satisfaction relation $\sigma, i \models I \in
\Delta_{\theta}$ is then defined as in $\tltll$.

Once again, since $\Delta_{\theta}$ behaves like an input determined
operator, each $\rtltl{\Sigma, \Rop}$ formula is also a $\tltl{\Sigma,
\Op}$ formula, for an appropriately chosen set of input determined
operators $\Op$, containing operators of the form $\Delta_{\theta}$.
A $\rtltl{\Sigma, \Rop}$ formula $\theta$ naturally defines both 
a timed language
$L(\theta) = \{ \sigma \in \two{\Sigma} \ | \ \sigma, 0 \models \theta
\}$ and 
a floating language $L^f(\theta) = 
\{(\sigma, i) \ | \ \sigma, i \models \theta \}$.

As an example, the formula $a \And ([1,1] \in \F_{\theta})$ where
$\theta = b \And \Ominus a \And \O a$, 
restates the property expressed by the $\rtmsol$ formula in
Sec.~\ref{section:mso2}.

Let us denote by $\rtfo{\Sigma, \Rop}$ the first-order fragment of the
logic $\rtmso{\Sigma, \Rop}$.
Then we have the following expressive completeness result:

\begin{theorem}
\label{thm:ltl2}
$\rtltl{\Sigma, \Rop}$ is expressively equivalent to
$\rtfo{\Sigma, \Rop}$.
\end{theorem}

\begin{proof}
As before we show that formulas in the logics are equivalent
level-wise (the level of a $\rtltll$ formula 
being defined analogous to $\rtmsol$).
We show by induction on $i$ that
\begin{enumerate}
\item
\label{item:ltl2:1}
A timed language $L \subseteq \two{\Sigma}$ is definable by a level
$i$ $\rtltl{\Sigma, \Rop}$ formula iff it is definable by a 
level $i$ $\rtfo{\Sigma, \Rop}$ sentence.
\item
\label{item:ltl2:2}
A floating timed language over $\Sigma$ is definable by a level $i$
$\rtltl{\Sigma, \Rop}$ formula iff it is definable by a level $i$
$\rtfo{\Sigma, \Rop}$ formula with one free variable.
\end{enumerate}

The base case for part~\ref{item:ltl2:1} follows from
Theorem~\ref{thm:kamp1}, since level 0 formulas are simply untimed
$\ltl{\Sigma}$ and $\fo{\Sigma}$ formulas.
For the base case for part~\ref{item:ltl2:2}, a level 0
$\rtltl{\Sigma, \Rop}$ formula $\theta$ can be translated to a level 0
$\rtfo{\Sigma, \Rop}$ formula $\psi$ with one free variable $z$ using
the translation given in the proof of Theorem~\ref{thm:ltl1}.
The formula $\psi$ satisfies $\sigma, [i/z] \models \psi$ iff 
$\sigma, i \models \theta$.
The converse direction follows immediately from the
following version of Kamp's result:
\begin{theorem}[\cite{k68}]
\label{thm:kamp2}
For any $\fo{A}$ formula $\psi$ with one free variable $z$, 
there is a $\ltl{A}$ formula $\theta$ s.t. for each $\alpha \in
A^{\omega}$ and $i \in \nat$, $\alpha, [i/z] \models
\psi$ iff $\alpha, i \models \theta$.
\end{theorem}

Turning now to the induction step, let $\theta$ be a level $i+1$
$\rtltl{\Sigma, \Rop}$ formula.
Let $\Op$ be the set of top-level $\Delta$ operators used in $\theta$.
Then $\theta$ is a $\tltl{\Sigma, \Op}$ formula, and hence by
Theorem~\ref{thm:ltl1} we have an equivalent $\tfo{\Sigma, \Op}$
sentence $\varphi$ (i.e. with $L(\theta) = L(\varphi)$).
Now each operator in $\Op$ is of the form $\Delta_{\eta}$ where $\eta$
is a level $i$ or less $\rtltl{\Sigma, \Rop}$ formula, and hence by
the induction hypothesis we have an equivalent $\rtfo{\Sigma, \Rop}$
formula $\psi$ with one free variable, such that $L^f(\eta) =
L^f(\psi)$.
It now follows that the input determined operators $\Delta_{\eta}$ and
$\Delta_{\psi}$ are semantically equivalent, and hence we can replace
each $\Delta_{\eta}$ by $\Delta_{\psi}$ in $\varphi$ to get an equivalent
$\rtfo{\Sigma, \Rop}$ sentence $\varphi'$.
By construction, the sentence $\varphi'$ is also of level $i+1$.
The converse direction is argued in a very similar manner, once again
factoring through Theorem~\ref{thm:ltl1}.

For part~\ref{item:ltl2:2}, a level $i+1$ $\rtltl{\Sigma, \Rop}$ 
formula $\theta$ is a $\tltl{\Sigma, \Op}$ formula, for the set of
operators $\Op$ defined above.
Now using the translation given in the proof of Theorem~\ref{thm:ltl1}
we obtain a $\tfo{\Sigma, \Op}$ formula $\psi$ with a one free
variable, satisfying $L^f(\theta) = L^f(\psi)$.
Again, by the induction hypothesis, we can replace each
$\Delta_{\eta}$ in $\Op$ with an equivalent $\Delta_{\phi}$, to get an
equivalent $\rtfo{\Sigma, \Rop}$ with the required properties.

In the converse direction, let $\psi$ be a level $i+1$ 
$\rtfo{\Sigma, \Rop}$ formula with one free variable $z$.
Let $\Op$ be set of top-level $\Delta$ operators in $\psi$.
Then $\psi$ is also a formula in $\tfo{\Sigma, \Op}$.
Let $\Gamma$ be the proper symbolic alphabet induced by $\psi$.
Then we can use the translation $\ttosfn$
(cf. Sec~\ref{section:tmso}) on $\psi$ (w.r.t. $\Gamma$) to get a
formula $\widehat{\psi}$ in $\fo{\Gamma}$ with one free variable
$z$ which preserves timed models.
By Kamp's theorem above, we have an equivalent $\ltl{\Gamma}$ formula
$\widehat{\theta}$
which preserves the floating language accepted.
Finally we can apply the translation $\stotfn$ on $\widehat{\theta}$
to get a $\tltl{\Sigma, \Op}$ formula $\theta$ which preserves timed
models (cf. Sec.~\ref{section:tltl}).
The formula $\theta$ satisfies the property that $L^f(\theta) =
L^f(\psi)$.

Now using the induction hypothesis each operator $\Delta_{\phi}$ in
$\theta$ can be replaced
 by an equivalent $\Delta_{\eta}$ operator, with
$\eta$ a $\tltl{\Sigma, \Op}$ formula,
to get an equivalent level $i+1$ $\rtltl{\Sigma, \Rop}$ formula
$\theta'$.
This ends the proof of Theorem~\ref{thm:ltl2}.
\qed
\end{proof}

\section{Expressive completeness of $\mitll$}
\label{section:mitl}

As an application of the results in this paper we
show that the logic $\mitll$ introduced
in \cite{afh96} is expressively equivalent to $\rtfol$ for a
suitably defined set of recursive input determined operators.
We point out here that this result is shown for the pointwise
semantics of $\mitll$ given below.
We begin with the logic $\mitle{\Sigma}$ which has the following
syntax \cite{ah93}:
\[
\theta ::= \ a \ | \
\O \theta \ | \ 
\Ominus \theta \ | \ 
(\theta \U_I \theta) \ | \
(\theta \S_I \theta) \ | \
\notof \theta \ | \
(\theta \Or \theta).
\]
Here $I$ is an interval in $\intervals$.
When $I$ is restricted to be \emph{non-singular} (i.e. not of the
form $[r,r]$) then we get the logic $\mitl{\Sigma}$.
The logic is interpreted over timed words in $\two{\Sigma}$
similarly to $\tltll$.
The modalities $\U_I$ and $\S_I$ are interpreted as follows, for a
timed word $\sigma = (\alpha, \tau)$.
\[
\begin{array}{lll}
\sigma, i \models \theta \U_I \eta & \mathrm{iff} & 
       \exists k \geq i:\, \sigma, k \models \eta, 
       \tau(k) - \tau(i) \in I \mathrm, {\ and \ }
\forall j:\, i \leq j < k, \ \sigma, j \models \theta \\
\sigma, i \models \theta \S_I \eta & \mathrm{iff} &
        \exists k \leq i:\, \sigma, k \models \eta,
        \tau(i) - \tau(k) \in I, \mathrm{\ and \ }
\forall j:\, k < j \leq i, \ \sigma, j \models \theta.
\end{array}
\]

We first observe that $\mitle{\Sigma}$ is expressively equivalent
to its sublogic $\mitlde{\Sigma}$ in which the modalities $U_I$
and $S_I$ are replaced by the modalities $U$, $S$, $\Fm_I$ and
$\Fpm_I$, where $U$ and $S$ are as usual and 
$\Fm_I \theta = \true \U_I \theta$ and $\Fpm_I \theta = \true
\S_I \theta$.
This is because the formula $\theta \U_{I} \eta$ (and dually
$\theta S_I \eta$) can be translated as follows.
Here `$\rangle$' denotes either a `$]$' or `$)$' interval bracket.
\[
\begin{array}{lcl}
\theta \U_{I} \eta & = & \left \{
		\begin{array}{ll}
	 \Fm_{I} \eta \And \G_{[0,a)}(\theta \U
              (\theta \And \O{\eta})) & \mathrm{if \ } I = [a,b\rangle, a>0 \\
         \Fm_{I} \eta \And \G_{[0,a]}(\theta \U
              (\theta \And \O{\eta})) & \mathrm{if \ } I = (a,b\rangle, a>0 \\
         \Fm_{I} \eta \And (\theta \U \eta)
                                       & \mathrm{if \ }I = [0,b\rangle \\
         \Fm_{I} \eta \And (\theta \U (\theta
              \And \O{\eta})) & \mathrm{if \ }I = (0,b\rangle.
	        \end{array}
	                  \right.
\end{array}
\]

Next we consider the logic
$\rtltl{\Sigma, \{\F,\Fp\}}$ where the semantics of the
recursive input
determined operators $\F$ and $\Fp$ are given below (as usual
$\sigma \in \two{\Sigma}$ with $\sigma = (\alpha, \tau)$).
\[
\begin{array}{lcl}
\sem{\F}(X,\sigma, i) & = \{ I \in \intervals \ | \ \exists j \in
X:\ j \geq i, \mathrm{\ and \ } \tau_j - \tau_i \in I \}\\
\sem{\Fp}(X,\sigma, i) & = \{ I \in \intervals \ | \ \exists j \in
X:\ j \leq i, \mathrm{\ and \ } \tau_i - \tau_j \in I \}.
\end{array}
\]
The logic $\mitlde{\Sigma}$ is clearly expressively equivalent
to $\rtltl{\Sigma, \{\F, \Fp\}}$ since the predicates $\Fm_{I} \theta$
and $I \in \F_{\theta}$ are equivalent.
Using Theorem~\ref{thm:ltl2} we can now conclude that
\begin{theorem}
\label{thm:mitl}
$\mitle{\Sigma}$ is expressively equivalent to $\rtfo{\Sigma,
    \{\F, \Fp \}}$.
\end{theorem}

Let $\rtfol_{\neq}$ denote the restriction of $\rtfol$ to
non-singular intervals.
Then since the translation of $\mitlel$ to $\mitldel$ does not
introduce any singular intervals, and the constructions in
Theorem~\ref{thm:ltl2} preserve the interval vocabulary of the
formulas, we conclude that the logics
$\mitl{\Sigma}$ and $\rtfol_{\neq}(\Sigma, \{\F, \Fp\})$ are
expressively equivalent.

\end{document}